\documentclass[prc,preprintnumbers,onecolumn]{revtex4}
\usepackage{graphicx}
\usepackage{amssymb}
\usepackage{amsmath}

\begin{document}

\title{Fusion at near-barrier energies within quantum diffusion approach}
\author{V.V.Sargsyan$^1$, G.G.Adamian$^{1}$, N.V.Antonenko$^1$, W. Scheid$^2$, and  H.Q.Zhang$^3$
}
\affiliation{$^{1}$Joint Institute for Nuclear Research, 141980 Dubna, Russia\\
$^{2}$Institut f\"ur Theoretische Physik der
Justus--Liebig--Universit\"at,
D--35392 Giessen, Germany\\
$^{3}$China Institute of Atomic Energy, Post Office Box 275, Beijing 102413,  China
}
\date{\today}


\begin{abstract}
Within the quantum diffusion approach  the role of neutron transfer in the fusion (capture)
reactions with  toughly and weakly bound nuclei is discussed. The breakup process is analyzed.
New methods for the study of  the breakup probability are suggested.
\end{abstract}

\maketitle

\section{Introduction}
\label{intro}
The  nuclear deformation and neutron-transfer process
have been identified as playing a major role in the
magnitude of the sub-barrier fusion (capture)  cross sections~\cite{Gomes}.
There are a several experimental evidences which confirm the importance
of  nuclear deformation on the  fusion.
The influence of nuclear deformation is straightforward.
If the target nucleus is prolate in the ground state,
the Coulomb field on its tips is lower than on its sides, that
then increases the capture or fusion probability at  energies below
the barrier corresponding to the spherical nuclei.
The role of neutron transfer
reactions is less clear.
%A correlation between the overall transfer
%strength and fusion enhancement was firstly noticed  in Ref.~\cite{Henning}.
The importance of neutron transfer with positive $Q$-values
on nuclear fusion (capture) originates from
the fact that neutrons are insensitive to the Coulomb barrier and
therefore they can start being transferred at larger separations
before the projectile is captured by target-nucleus.
%~\cite{Stelson}.
Therefore, it is generally thought
that  the sub-barrier
fusion cross section will increase
%~\cite{Pengo,Roberts,Stefanini3236s110pd,Sonzogni}
because of the neutron transfer.

The fusion (capture) dynamics induced by loosely bound radioactive
ion beams  is currently being extensively
studied.
However, the long-standing  question whether fusion (capture) is enhanced
or suppressed with these beams has not yet been answered unambiguously.
The study of the fusion reactions involving nuclei at the drip-lines has
led to contradictory results.

\section{Quantum diffusion approach for capture}
\label{sec-1}
In the quantum diffusion approach~\cite{EPJSub,EPJSub02,EPJSub1,EPJSub2,EPJSub3}
the capture of the projectile  by the target-nucleus is described with
a single relevant collective variable: the relative distance  between
the colliding nuclei. This approach takes into consideration the fluctuation and dissipation effects in
collisions of heavy ions which model the coupling of the relative motion with various channels
(for example, the non-collective single-particle excitations, low-lying collective dynamical modes
 of the target and projectile).
The  nuclear static deformation effects
are taken into account through the dependence of the nucleus-nucleus potential
on the deformations and mutual orientations of the colliding nuclei.
We have to mention that many quantum-mechanical and non-Markovian effects accompanying
the passage through the potential barrier are taken into consideration in our
formalism~\cite{EPJSub,EPJSub1}.

The capture cross section is a sum of partial capture cross sections~\cite{EPJSub,EPJSub1}
\begin{eqnarray}
\sigma_{cap}(E_{\rm c.m.})&=&\sum_{J}^{}\sigma_{\rm cap}(E_{\rm
c.m.},J)=\nonumber\\&=& \pi\lambdabar^2
\sum_{J}^{}(2J+1)\int_0^{\pi/2}d\theta_1\sin(\theta_1)\int_0^{\pi/2}d\theta_2\sin(\theta_2) P_{\rm cap}(E_{\rm
c.m.},J,\theta_1,\theta_2),
\label{1a_eq}
\end{eqnarray}
where $\lambdabar^2=\hbar^2/(2\mu E_{\rm c.m.})$ is the reduced de Broglie wavelength,
$\mu=m_0A_1A_2/(A_1+A_2)$ is the reduced mass ($m_0$ is the nucleon mass),
and the summation is over the possible values of angular momentum $J$
at a given bombarding energy $E_{\rm c.m.}$.
Knowing the potential of the interacting nuclei for each orientation with
the angles $\theta_i (i=1,2)$, one can obtain the partial capture probability
$P_{\rm cap}$ which is defined by the passing probability of the potential barrier
in the relative distance $R$ coordinate at a given $J$.
The value of $P_{\rm cap}$
is obtained by integrating the propagator $G$ from the initial
state $(R_0,P_0)$ at time $t=0$ to the final state $(R,P)$ at time $t$ ($P$ is a momentum):
\begin{eqnarray}
P_{\rm cap}=\lim_{t\to\infty}\int_{-\infty}^{r_{\rm in}}dR\int_{-\infty}^{\infty}dP\
G(R,P,t|R_0,P_0,0)
%\nonumber \\
=\lim_{t\to\infty}\frac{1}{2} {\rm erfc}\left[\frac{-r_{\rm in}+\overline{R(t)}}
{{\sqrt{\Sigma_{RR}(t)}}}\right].
\label{1ab_eq}
\end{eqnarray}
The second line in (\ref{1ab_eq}) is obtained by using the propagator
$G=\pi^{-1}|\det {\bf \Sigma}^{-1}|^{1/2}
\exp(-{\bf q}^{T}{\bf \Sigma}^{-1}{\bf q})$
(${\bf q}^{T}=[q_R,q_P]$,
$q_R(t)=R-\overline{R(t)}$, $q_P(t)=P-\overline{P(t)}$, $\overline{R(t=0)}=R_0$,
$\overline{P(t=0)}=P_0$, $\Sigma_{kk'}(t)=2\overline{q_k(t)q_{k'}(t)}$, $\Sigma_{kk'}(t=0)=0$,
$k,k'=R,P$) calculated  for
an inverted oscillator which approximates
the nucleus-nucleus potential $V$ in the variable $R$.
The frequency $\omega$ of
this oscillator with an internal turning point $r_{\rm in}$ is defined from the condition of equality of the
classical actions of approximated and realistic potential barriers of the same hight at given $J$.
%It should be noted that the passage through the Coulomb barrier approximated by a parabola
%has been previously studied in Refs.~\cite{EPJSub,EPJSub1,Hofman}.
This approximation is well justified for the
reactions and energy range, which are here considered.

We assume that the sub-barrier capture  mainly  depends  on the optimal one-neutron ($Q_{1n}>Q_{2n}$) or
two-neutron ($Q_{2n}>Q_{1n}$) transfer with the  positive  $Q$-value.
% among all possible transfer channels.
Our assumption is that, just before the projectile is captured by the target-nucleus
(just before the crossing of the Coulomb barrier) which is a slow process,
the  transfer  occurs   and can lead to the
population of the first excited collective state in the recipient nucleus~\cite{SSzilner}
(the donor nucleus remains in the ground state).
So, the motion to the
$N/Z$ equilibrium starts in the system before the capture
because it is energetically favorable in the dinuclear system in the vicinity of the Coulomb barrier.
For the reactions under consideration,
the average change of mass asymmetry is connected to the one- or two-neutron
transfer ($1n$- or $2n$-transfer).
Since after the transfer the mass numbers, the isotopic composition and the deformation parameters
of the interacting nuclei, and, correspondingly, the height $V_b=V(R_b)$
%($R_b$ is the position of the Coulomb barrier) are changed,
and shape of the Coulomb barrier are changed,
one can expect an enhancement or suppression of the capture.
If  after the neutron transfer the deformations of interacting nuclei increase (decrease),
the capture probability increases (decreases).
When the isotopic dependence of the nucleus-nucleus
potential is weak and   after the transfer the deformations of interacting nuclei do not change,
there is no effect of the neutron transfer on the capture.
In comparison with Ref.~\cite{Dasso}, we assume that the negative transfer $Q-$values
do not play  visible  role in the capture process.
Our scenario was verified in the description of many reactions~\cite{EPJSub1,EPJSub2,EPJSub3}.

\section{Results of calculations}
\label{sec-2}
Because the capture cross section is equal to the complete fusion cross section for the reactions treated,
the quantum diffusion approach for the capture is applied to study the complete fusion.
All calculated results are obtained with the same set of parameters
as in Ref.~\cite{EPJSub}.
% and are rather insensitive to the reasonable variation of them~\cite{EPJSub,EPJSub1}.
Realistic friction coefficient in the relative distance coordinate
$\hbar\lambda$=2 MeV
is
used.  Its value
is close to that calculated within the mean-field approaches~\cite{obzor}.
%The value of $\tilde\lambda$ is set to obtain this value of $\hbar\lambda$.
%The heights of the calculated Coulomb barriers $V_b=V(R_b)$
%($R_b$ is the position of the Coulomb barrier)
%are adjusted to the experimental data for the fusion or capture cross sections.
For the nuclear part of the nucleus-nucleus
potential, the double-folding formalism with
the Skyrme-type density-dependent effective
nucleon-nucleon interaction is used~\cite{EPJSub,EPJSub1}.
The parameters of the nucleus-nucleus interaction potential $V(R)$
are adjusted to describe the experimental
data at energies above the Coulomb barrier corresponding to spherical nuclei.
The absolute values of the experimental quadrupole deformation parameters $\beta_2$
of even-even deformed nuclei in the ground state
and of the first excited collective states of nuclei
are taken from Ref.~\cite{Ram}.
For the  nuclei deformed in the
ground state, the $\beta_2$ in the first excited collective state is similar
to the $\beta_2$ in the ground state.
For the quadruple deformation parameter
of an odd nucleus, we choose the maximal value from the
deformation parameters of neighboring even-even nuclei (for example,
$\beta_2$($^{231}$Th)=$\beta_2$($^{233}$Th)=$\beta_2$($^{232}$Th)=0.261).
%In Ref.~\cite{Ram} the quadrupole
%deformation parameters $\beta_2$ are given for the first excited
%2$^{+}$ states of nuclei. For the  nuclei deformed in the
%ground state, the $\beta_2$ in 2$^{+}$ state is similar
%to the $\beta_2$ in the ground state and we use $\beta_2$
%from Ref.~\cite{Ram} in the calculations.
For the double magic and neighboring nuclei,
we take $\beta_2=0$ in the ground state.
Since there are  uncertainties in the definition of the values of $\beta_2$
in  light-mass nuclei,
one can extract the ground-state
quadrupole deformation parameters of
these  nuclei from a comparison
of the calculated capture cross sections with the existing experimental data.
% (Figs.~5-8) ($Q_{2n}$ are negative or close to zero).
%The best case is
%when the projectile or target is the spherical double magic nucleus and
%there are no neutron transfer channels with  positive $Q$-values.
%There are the uncertainties in the definition of the deformation of the light nucleus
%and we have no practically experimental deformation parameters for the odd-even, even-odd, and
%odd-odd nuclei.
By describing the   reactions
$^{12}$C+$^{208}$Pb,
$^{18}$O+$^{208}$Pb,
$^{32,36}$S+$^{90}$Zr,
$^{34}$S+$^{168}$Er,
$^{36}$S+$^{90,96}$Zr,
$^{58}$Ni + $^{58}$Ni,
and $^{64}$Ni + $^{58}$Ni,
where there are no neutron transfer
channels with positive $Q$-values,
we extract   the ground-state
quadrupole deformation parameters
$\beta_2$=-0.3,   0.1,  0.312,
0.1,   0,   0.05,  and  0.087,
for the nuclei $^{12}$C,
$^{18}$O, $^{32}$S,
$^{34}$S, $^{36}$S, $^{58}$Ni, and $^{64}$Ni, respectively,
which are used in our calculations.
%For the   $^{32}$S nucleus, the extracted $\beta_2$ is equal to the
%experimental one from Ref.~\cite{Ram}.

\subsection{Role of neutron transfer in capture process at  sub-barrier energies}
\label{sec-3}
After the neutron
transfer in the reaction
$^{40}$Ca($\beta_2=0$) + $^{96}$Zr($\beta_2=0.08$)$\to ^{42}$Ca($\beta_2=0.247$) + $^{94}$Zr($\beta_2=0.09$) (Fig.~1)
or
$^{40}$Ca($\beta_2=0$) + $^{124}$Sn($\beta_2=0.095$)$\to ^{42}$Ca($\beta_2=0.247$) + $^{122}$Sn($\beta_2=0.1$) (Fig.~1)
the deformation of the nuclei increases and the
mass asymmetry of the system decreases,  and,
thus, the value of the Coulomb barrier decreases and
the capture cross section becomes larger (Fig.~1).
In Fig.~2, we observe the same behavior in the reactions
$^{58}$Ni($\beta_2=0.05$) + $^{132}$Sn($\beta_2=0$)$\to ^{60}$Ni($\beta_2=0.207$) + $^{130}$Sn($\beta_2=0$)
($Q_{2n}=7.8$ MeV),
$^{58}$Ni($\beta_2=0.05$) + $^{130}$Te($\beta_2=0$)$\to ^{60}$Ni($\beta_2=0.207$) + $^{128}$Te($\beta_2=0$)
($Q_{2n}=5.9$ MeV),
$^{64}$Ni($\beta_2=0.087$) + $^{132}$Sn($\beta_2=0$)$\to ^{66}$Ni($\beta_2=0.158$) + $^{130}$Sn($\beta_2=0$)
($Q_{2n}=2.5$ MeV), and
$^{64}$Ni($\beta_2=0.087$) + $^{130}$Te($\beta_2=0$)$\to ^{66}$Ni($\beta_2=0.158$) + $^{128}$Te($\beta_2=0$)
($Q_{2n}=0.55$ MeV).
One can see a good agreement between the
calculated results and the experimental data~\cite{Liang,TimmersCa40Zr96,Stefanini40ca116124sn}.
So,
 the observed capture enhancement at  sub-barrier energies in
the reactions mentioned above
is related to the two-neutron transfer channel.
One can see that at energies above and near  the Coulomb barrier
the cross sections with and without
two-neutron transfer are almost similar.
% which was observed in Ref.~\cite{Liang}.
Since the two-neutron transfer causes a larger change of the deformations of the nuclei in
the reactions $^{58}$Ni + $^{132}$Sn,$^{130}$Te than  in the reactions
$^{64}$Ni + $^{132}$Sn,$^{130}$Te, at sub-barrier energies
the  capture enhancement  in the
reactions with $^{58}$Ni is larger than in the reactions with $^{64}$Ni  (Fig.~2).
%In Fig.~3 there are
%no significant differences  in the calculated reduced excitation
%functions $^{58,64}$Ni + $^{132}$Sn,$^{130}$Te at $E_{\rm c.m.}/V_b\approx\> 0.97$.
%Indeed, the same  was observed in Ref.~\cite{Liang}.
%A small differences between our reduced excitation
%functions and those of Ref.~\cite{Liang}
% appear because of different $V_b$ and $R_b$ of our model and Bass model
% used in~\cite{Liang}.
%We would expect a
%small  difference  in the reduced excitation
%functions of
%$^{40}$Ca + $^{96}$Zr,$^{124,132}$Sn at $E_{\rm c.m.}/V_b\approx\> 0.95$.
%
\begin{figure}[htb]
\centering
%\sidecaption
\includegraphics[scale=1]{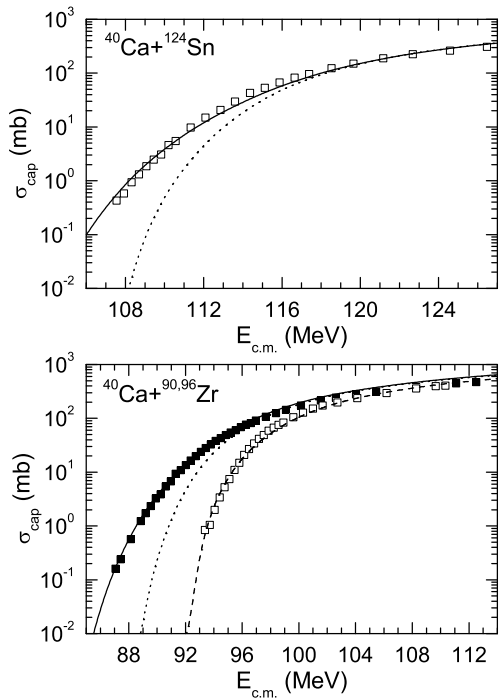}
\caption{
The calculated capture cross sections versus $E_{\rm c.m.}$ for the indicated reactions
$^{40}$Ca + $^{96}$Zr (solid line),  $^{40}$Ca + $^{90}$Zr (dashed line),
and $^{48}$Ca + $^{124}$Sn (solid line).
For the reactions $^{40}$Ca + $^{96}$Zr,$^{124}$Sn,
the calculated capture cross sections without
the neutron transfer process are shown by dotted lines.
The experimental data (symbols)
are from Refs.~\protect\cite{TimmersCa40Zr96,Stefanini40ca116124sn}.
}
\label{1_fig}
\end{figure}
\begin{figure}[htb]
\centering
%\sidecaption
\includegraphics[scale=1]{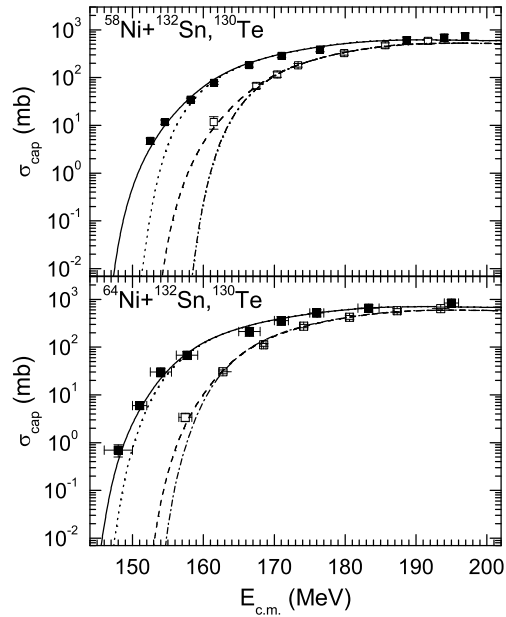}
\caption{
The same as in Fig.~1,  for the  reactions
$^{58,64}$Ni + $^{132}$Sn   (solid lines) and $^{58,64}$Ni + $^{130}$Te (dashed lines).
The experimental  data  (symbols) are  from Refs.~\protect\cite{Liang,Liang0}.
For the reactions $^{58,64}$Ni + $^{132}$Sn   (dotted lines)
and $^{58,64}$Ni + $^{130}$Te (dash-dotted lines),
the calculated capture cross sections without the
 neutron transfer  are shown.
}
\label{2_fig}
\end{figure}
%\begin{figure}[htb]
%\centering
%%\sidecaption
%\includegraphics[scale=1]{Figure3.eps}
%\caption{
%The calculated reduced capture cross sections versus
%$E_{\rm c.m.}/V_b$ for the indicated reactions.
%The results with and without taking into account
%the two-neutron transfer process are shown by  solid  and dotted lines, respectively.
%Here, $V_b=V(R_b)$ and $R_b$   are the heights and positions, respectively,
%of the calculated Coulomb barriers.
%}
%\label{3_fig}
%\end{figure}

%---------------------------------------------------
%
\begin{figure}[htb]
\centering
%\sidecaption
\includegraphics[scale=1]{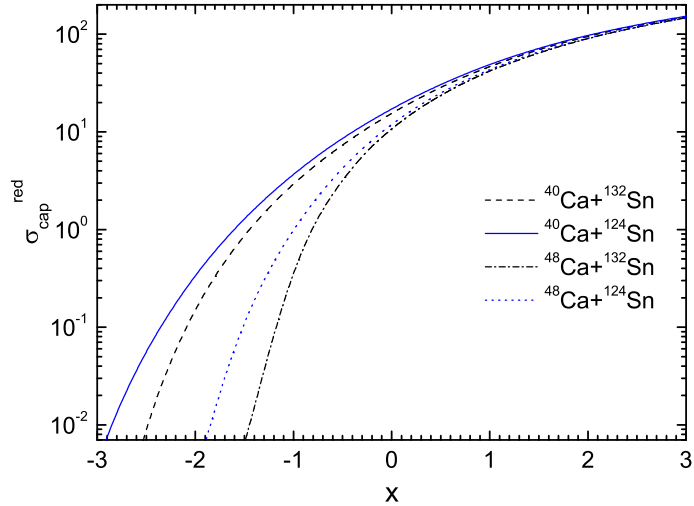}
\caption{
(Color online) The calculated reduced capture cross sections versus $(E_{\rm c.m.}-V_b)/(\hbar\omega_b)$
in the reactions
$^{40}$Ca+$^{124}$Sn (solid line),
$^{48}$Ca+$^{124}$Sn (dashed line), $^{48}$Ca+$^{124}$Sn (dotted line),
and  $^{48}$Ca+$^{132}$Sn (dash-dotted line).
}
\label{3_fig}
\end{figure}
\begin{figure}[htb]
\centering
%\sidecaption
\includegraphics[scale=1]{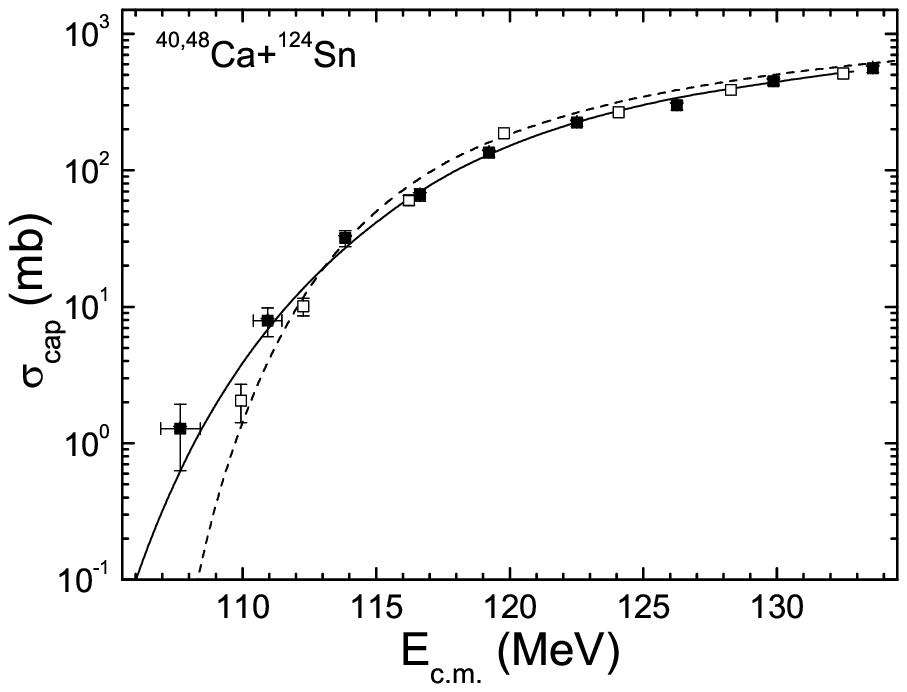}
\caption{
(Color online)
The calculated  capture cross sections versus $E_{\rm c.m.}$ for the reactions
$^{40}$Ca+$^{124}$Sn (solid line) and $^{48}$Ca+$^{124}$Sn (dashed line).
The experimental data  for the reactions $^{40}$Ca+$^{124}$Sn (solid squares)
and $^{48}$Ca+$^{124}$Sn (open squares) are from Ref.~\protect\cite{Kol}.
In the calculations the barriers were adjusted to the experimental values.
}
\label{4_fig}
\end{figure}
\begin{figure}[htb]
\centering
%\sidecaption
\includegraphics[scale=1]{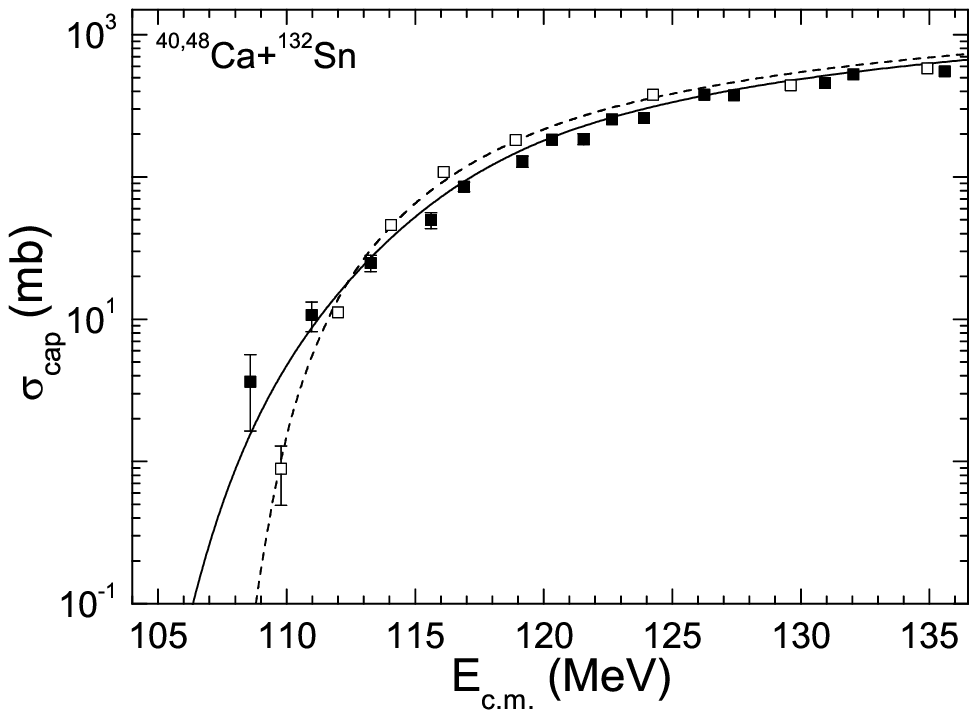}
\caption{
(Color online)
The calculated  capture cross sections versus $E_{\rm c.m.}$ for the reactions
$^{40}$Ca+$^{132}$Sn (solid line) and $^{48}$Ca+$^{132}$Sn (dashed line).
The experimental data  for the reactions $^{40}$Ca+$^{132}$Sn (solid squares) and $^{48}$Ca+$^{132}$Sn (open squares)
are from Ref.~\protect\cite{Kol}.
In the calculations the barriers were adjusted to the experimental values.
}
\label{5_fig}
\end{figure}
%
%\begin{figure}[htb]
%\centering
%%\sidecaption
%\includegraphics[scale=1]{Figure14.eps}
%\caption{
%The calculated  capture cross sections vs
%$(E_{\rm c.m.}-V_b)$ for the  reactions $^{48}$Ca+$^{126}$Sn (solid line) (a),
%$^{40}$Ca+$^{126}$Sn (dashed line) (a), $^{36}$S+$^{132}$Sn (solid line) (b), and
%$^{32}$S+$^{132}$Sn  (dashed line) (b).
%The results  without taking into account
%the two-neutron transfer process are shown by   dotted lines for the  reactions $^{40}$Ca+$^{126}$Sn (a) and
% $^{32}$S+$^{132}$Sn (b).
%}
%\label{14_fig}
%\end{figure}
%
One can  make unambiguous statements
regarding the neutron transfer process with a positive $Q$-value when the colliding nuclei
are  double magic or semi-magic.
In this case one can disregard the  deformation and orientation effects
before the neutron transfer. To eliminate the influence
of the nucleus-nucleus
potential on the capture (fusion) cross section and to make conclusions about the role of deformation of
colliding nuclei and the nucleon transfer between interacting nuclei in the capture  (fusion) cross
section,  a reduction procedure is useful~\cite{Gomes2}.
It consists of the following transformations:
$$E_{\rm c.m.} \rightarrow x= \dfrac{E_{\rm c.m.}-V_b}{\hbar \omega_b},
\qquad \sigma_{cap} \rightarrow \sigma_{cap}^{red}=\dfrac{2 E_{\rm c.m.}}{\hbar \omega_b R_b^{2}}\sigma_{cap},$$
where  $\sigma_{cap}=\sigma_{cap}(E_{\rm c.m.})$ is the capture cross section at bombarding energy
$E_{\rm c.m.}$.
The frequency $\omega_b =\sqrt{V^{''}(R_b)/\mu}$
is related with the second derivative $V^{''}(R_b)$ of the total nucleus-nucleus potential $V(R)$
(the Coulomb + nuclear parts)
at the barrier position $R_b$.
With these replacements we  compared  the  reduced calculated capture (fusion) cross sections
$\sigma_{cap}^{red}$ for the reactions $^{40,48}$Ca+$^{124,132}$Sn (Fig.~3).
The choice of the projectile-target combination is crucial, and for the systems
studied one can  make unambiguous statements
regarding the neutron transfer process with a positive $Q$-value when the interacting nuclei
are  double magic or semi-magic spherical nuclei.
In this case one can disregard the strong  direct nuclear deformation effects.
In Fig.~3,
one can see that the reduced capture cross sections in the  reactions $^{40}$Ca+$^{124,132}$Sn
with the positive $Q_{2n}$-values
strongly deviate from   those in the reactions
$^{48}$Ca+$^{124,132}$Sn,
where the neutron transfers   are suppressed because of the negative $Q$-values.
 After two-neutron transfer in the reactions
 $^{40}$Ca($\beta_2=0$)+$^{124}$Sn($\beta_2=0.1$)$\to ^{42}$Ca($\beta_2=0.25$)+$^{122}$Sn($\beta_2=0.1$)
 ($Q_{2n}$=5.4 MeV)
 and
$^{40}$Ca($\beta_2=0$)+$^{132}$Sn($\beta_2=0$)$\to ^{42}$Ca($\beta_2=0.25$)+$^{130}$Sn($\beta_2=0$)
($Q_{2n}$=7.3 MeV)
the deformation of the light nucleus increases and the mass asymmetry of the system decreases  and,
thus, the value of the Coulomb barrier decreases and
the capture cross section becomes larger (Fig.~3). So, because of the transfer effect
the systems $^{40}$Ca+$^{124,132}$Sn show large sub-barrier enhancements with respect to
the systems $^{48}$Ca+$^{124,132}$Sn.
We observe that the $\sigma_{cap}^{red}$ in the  $^{40}$Ca+$^{124}$Sn ($^{48}$Ca+$^{124}$Sn)
reaction are larger
than those in the  $^{40}$Ca+$^{132}$Sn  ($^{48}$Ca+$^{132}$Sn)   reaction. The reason of that
is the nonzero quadrupole deformation  of the heavy nucleus $^{124}$Sn.
It should be stressed that there are almost no difference
between  $\sigma_{cap}^{red}$ in the reactions $^{40,48}$Ca+$^{124,132}$Sn
 at energies above the Coulomb barrier.

In Figs.~4 and 5 one can see a good agreement between the  calculated results and the experimental data
in the reactions $^{40,48}$Ca+$^{124,132}$Sn.
This means that the observed capture enhancements  in
the reactions $^{40}$Ca+$^{124,132}$Sn   at sub-barrier energies
are related to the two-neutron transfer effect.
Note that the slope of the excitation function strongly depends on the deformations of the
interacting nuclei and, respectively, on the neutron transfer effect.

To describe the reactions  $^{40,48}$Ca+$^{132}$Sn  and  $^{48}$Ca+$^{124,132}$Sn (Figs. 4 and 5),
we extracted the values of the corresponding  Coulomb barrier $V_b$ for the spherical nuclei.
There are  differences between  the calculated and extracted $V_b$.
From the direct  calculations
of the nucleus-nucleus potentials (with the same set of parameters), we obtained
$V_b$($^{40}$Ca+$^{124}$Sn)-$V_b$($^{48}$Ca+$^{124}$Sn)=2.3 MeV,
$V_b$($^{40}$Ca+$^{132}$Sn)-$V_b$($^{48}$Ca+$^{132}$Sn)=2.2 MeV,
$V_b$($^{40}$Ca+$^{124}$Sn)-$V_b$($^{40}$Ca+$^{132}$Sn)=1.3 MeV,
and
$V_b$($^{48}$Ca+$^{124}$Sn)-$V_b$($^{48}$Ca+$^{132}$Sn)=1.2 MeV.
From the extractions, we got
$V_b$($^{40}$Ca+$^{124}$Sn)-$V_b$($^{48}$Ca+$^{124}$Sn)=1.1 MeV,
$V_b$($^{40}$Ca+$^{132}$Sn)-$V_b$($^{48}$Ca+$^{132}$Sn)=1.0 MeV,
$V_b$($^{40}$Ca+$^{124}$Sn)-$V_b$($^{40}$Ca+$^{132}$Sn)=-0.3 MeV,
and
$V_b$($^{48}$Ca+$^{124}$Sn)-$V_b$($^{48}$Ca+$^{132}$Sn)=-0.4 MeV,
which seem to be  unrealistically small.
%It could be some problem in the experimental energy calibration.
However,  these differences of $V_b$
%about of 1.2 MeV
do not influence the slopes of the excitation functions but only
lead to the shifting of the energy scale.
With realistic isospin trend of $V_b$
$\sigma_{cap}$($^{40}$Ca+$^{124}$Sn)$< \sigma_{cap}$($^{48}$Ca+$^{124}$Sn)
and
$\sigma_{cap}$($^{40}$Ca+$^{132}$Sn)$< \sigma_{cap}$($^{48}$Ca+$^{132}$Sn)
at energies above the corresponding Coulomb barriers.

\begin{figure}[htb]
\centering
%\sidecaption
\includegraphics[scale=1]{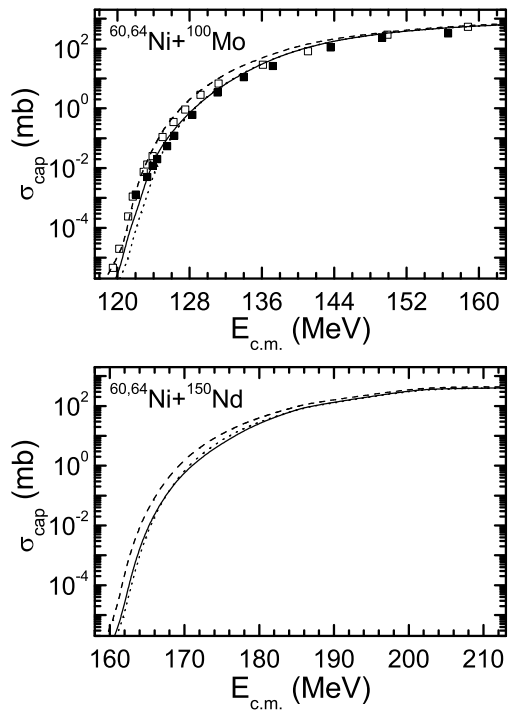}
\caption{
(Color online)
The same as in Fig.~1,  for the  indicated  reactions
$^{60}$Ni + $^{100}$Mo,$^{150}$Nd (solid  lines),
and  $^{64}$Ni + $^{100}$Mo,$^{150}$Nd (dashed lines).
For the reactions $^{60}$Ni + $^{100}$Mo
and $^{60}$Ni + $^{150}$Nd,
the calculated capture cross sections without the
neutron transfer  are shown by dotted lines.
The experimental  data  for the  reactions
$^{60}$Ni + $^{100}$Mo (closed squares) and $^{64}$Ni + $^{100}$Mo (open squares)
are  from
Ref.~\protect\cite{Scarlassara}.
}
\label{6_fig}
\end{figure}
One can find  reactions with a positive $Q$-values of the two-neutron
transfer  where the transfer weakly influences
or even suppresses the capture process. This happens if
after the transfer the deformations of the nuclei  do
not  change  much or even decrease. For instance, in the reactions
$^{60}$Ni($\beta_2\approx 0.1$) + $^{100}$Mo($\beta_2=0.231$)$\to ^{62}$Ni($\beta_2=0.198$) + $^{98}$Mo($\beta_2=0.168$)
($Q_{2n}=4.2$ MeV),
$^{64}$Ni($\beta_2\approx 0.087$) + $^{100}$Mo($\beta_2=0.231$)$\to ^{66}$Ni($\beta_2=0.158$) + $^{98}$Mo($\beta_2=0.168$)
($Q_{2n}=0.94$ MeV),
and
$^{60}$Ni($\beta_2\approx 0.1$) + $^{150}$Nd($\beta_2=0.285$)$\to ^{62}$Ni($\beta_2=0.198$) + $^{148}$Nd($\beta_2=0.204$)
($Q_{2n}=6$ MeV)
we expect a weak dependence of the capture cross section on the neutron transfer~(Fig.~6).
There is the experimental evidence  \cite{Scarlassara} of such an effect for the
 $^{60}$Ni + $^{100}$Mo reaction.
 So, the two-neutron transfer channel with large positive  $Q_{2n}$-value weakly influences
the fusion (capture) cross section.
The reduced capture cross sections in the reactions $^{60}$Ni + $^{100}$Mo,$^{150}$Nd
are close to each other in contrast to those in the reactions
$^{58,64}$Ni + $^{132}$Sn,$^{130}$Te.
The  $^{60}$Ni + $^{150}$Nd reaction has even a small suppression
due to the neutron transfer.

%---------------------------------------------------
\begin{figure}[htb]
\centering
%\sidecaption
\includegraphics[scale=1]{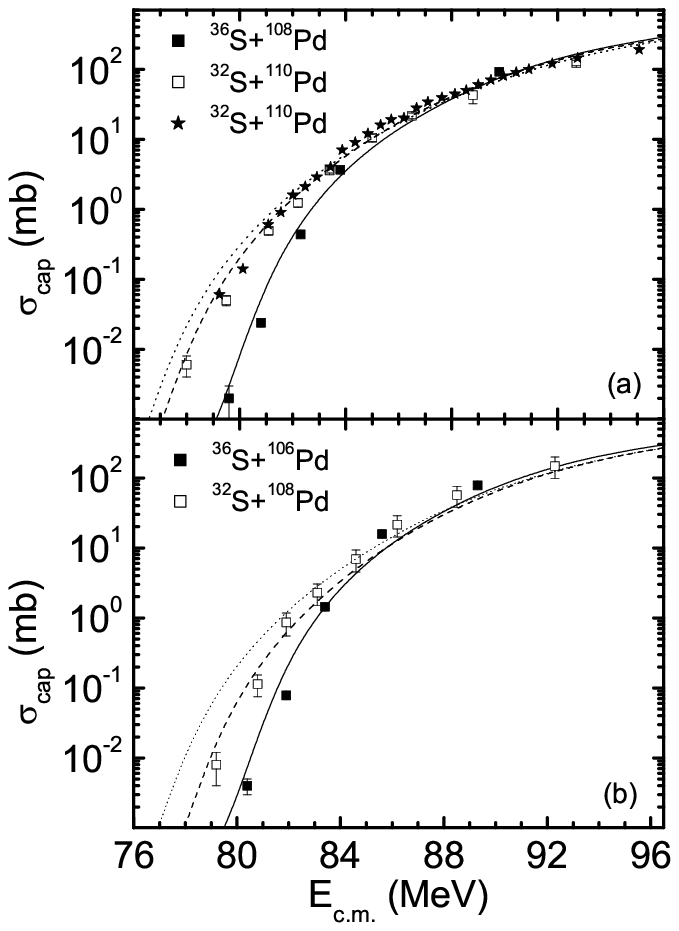}
\caption{
The calculated capture cross sections vs $E_{\rm c.m.}$ for the
reactions
$^{32}$S+$^{108,110}$Pd (dashed lines)
and $^{36}$S+$^{106,108}$Pd (solid lines) (a,b).
For the $^{32}$S+$^{110}$Pd reaction (a),
the calculated capture cross section without
the neutron transfer process is shown by a dotted line.
%The experimental data (symbols) are from Refs.~\protect\cite{Pengo,Stefanini3236s110pd}.
For the
reactions
$^{32}$S+$^{110}$Pd, the experimental data from~\protect\cite{Pengo}
and~\protect\cite{Stefanini3236s110pd} are marked by open squares and stars, respectively.
}
\label{7_fig}
\end{figure}
\begin{figure}[htb]
\centering
%\sidecaption
\includegraphics[scale=1]{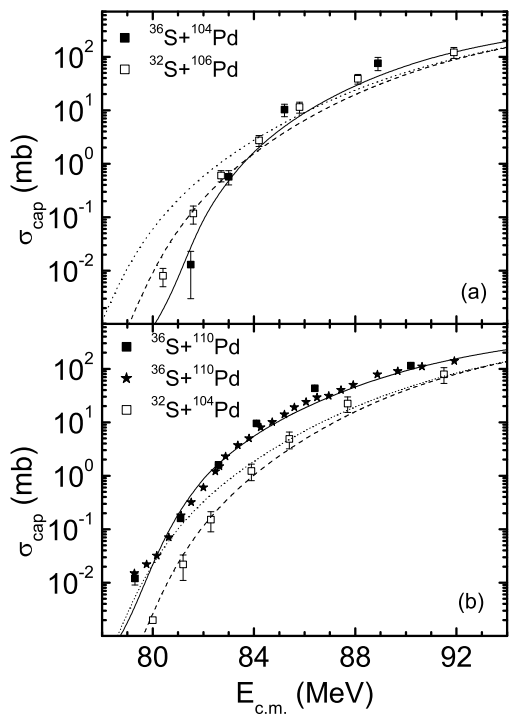}
\caption{
The same as in Fig.~7,  for the  reactions
$^{32}$S+$^{104,106}$Pd (dashed lines)
and $^{36}$S+$^{104,110}$Pd (solid lines) (a,b).
The dotted lines correspond to the reactions $^{32}$S+$^{104,106}$Pd
when the neutron transfer is disregarded.
The experimental data (symbols) are  from Ref.~\protect\cite{Pengo}.
}
\label{8_fig}
\end{figure}
\begin{figure}[htb]
\centering
%\sidecaption
\includegraphics[scale=1]{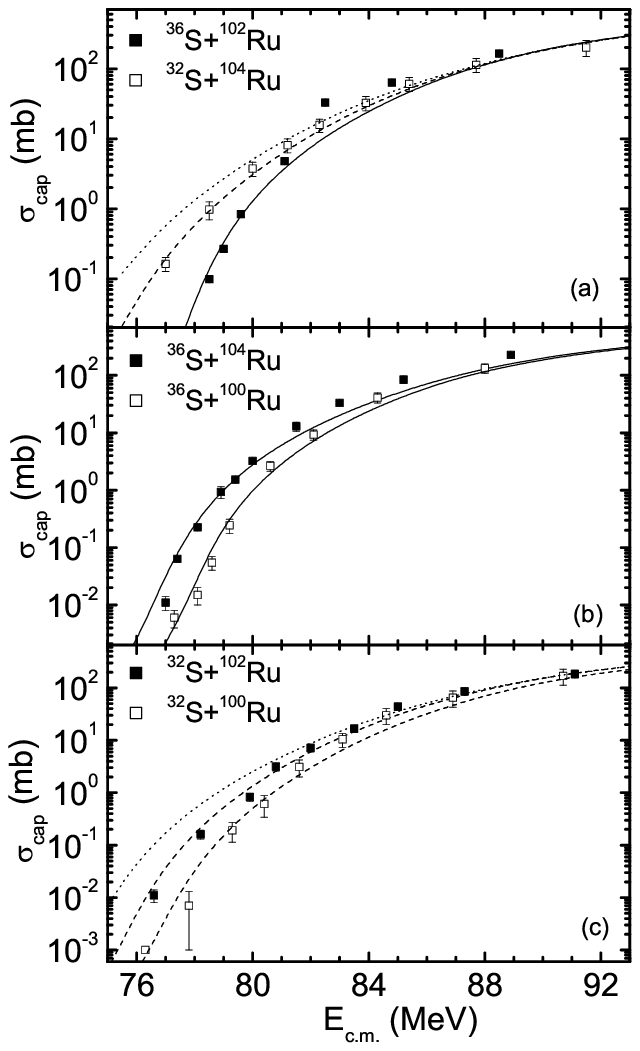}
\caption{
The calculated capture cross sections vs $E_{\rm c.m.}$
 for the  reactions
$^{32}$S+$^{100,102,104}$Ru (dashed lines) (a,b,c)
and
$^{36}$S+$^{100,102,104}$Ru (solid lines) (a,b). The dotted lines correspond to the reactions
$^{32}$S+$^{102,104}$Ru (a,c)
when the neutron transfer is disregarded.
The experimental data (symbols) are  from Ref.~\protect\cite{Pengo}.
}
\label{9_fig}
\end{figure}
%
%\begin{figure}[htb]
%\centering
%%\sidecaption
%\includegraphics[scale=1]{Figure10.eps}
%\caption{
%The same as in Fig.~8,  for the  reactions
%$^{32}$S+$^{98,100}$Mo (dashed lines)
%and $^{36}$S+$^{96,98}$Mo (solid lines) (a,b).
%The dotted lines correspond to the reactions $^{32}$S+$^{98,100}$Mo
%when the neutron transfer is disregarded.
%The experimental data (symbols) are  from Ref.~\protect\cite{Pengo}.
%}
%\label{10_fig}
%\end{figure}
%
%\begin{figure}[htb]
%\centering
%%\sidecaption
%\includegraphics[scale=1]{Figure11.eps}
%\caption{
%The calculated capture cross sections vs $E_{\rm c.m.}$  for the  reactions
%$^{32}$S+$^{92,94,96}$Mo (dashed lines)
%and $^{36}$S+$^{92,94,100}$Mo (solid lines) (a,b).
%The experimental data (symbols) are  from Ref.~\protect\cite{Pengo}.
%}
%\label{11_fig}
%\end{figure}
%
Figures  7-9 show the capture excitation function  for the reactions $^{32,36}$S+Pd,Ru
as a function of  the bombarding energy.
One can see a relatively good agreement between the
calculated results and the experimental data~\cite{Pengo}.
The $Q_{2n}$-values for the $2n$-transfer
processes are positive (negative) for all reactions with $^{32}$S ($^{36}$S).
%except $^{92}$Mo.
%For the  $^{32}$S+$^{92}$Mo ($^{36}$S+$^{92}$Mo) reaction, $Q_{2n}<0$ ($Q_{2n}>0$).
% (see Table 1).
At energies above and near  the Coulomb barrier
the cross sections with and without
two-neutron transfer are almost similar.
After the $2n$-transfer (before the capture)
in the reactions
$^{32}$S($\beta_2=0.312$)+$^{110}$Pd($\beta_2=0.257$)$\to ^{34}$S($\beta_2=0.252$)+$^{108}$Pd($\beta_2=0.243$),
$^{32}$S($\beta_2=0.312$)+$^{108}$Pd($\beta_2=0.243$)$\to ^{34}$S($\beta_2=0.252$)+$^{106}$Pd($\beta_2=0.229$),
$^{32}$S($\beta_2=0.312$)+$^{106}$Pd($\beta_2=0.229$)$\to ^{34}$S($\beta_2=0.252$)+$^{104}$Pd($\beta_2=0.209$),
$^{32}$S($\beta_2=0.312$)+$^{104}$Pd($\beta_2=0.209$)$\to ^{34}$S($\beta_2=0.252$)+$^{102}$Pd($\beta_2=0.196$),
or
$^{32}$S($\beta_2=0.312$)+$^{104}$Ru($\beta_2=0.271$)$\to ^{34}$S($\beta_2=0.252$)+$^{102}$Ru($\beta_2=0.24$),
$^{32}$S($\beta_2=0.312$)+$^{102}$Ru($\beta_2=0.24$)$\to ^{34}$S($\beta_2=0.252$)+$^{100}$Ru($\beta_2=0.215$),
$^{32}$S($\beta_2=0.312$)+$^{100}$Ru($\beta_2=0.215$)$\to ^{34}$S($\beta_2=0.252$)+$^{98}$Ru($\beta_2=0.195$)
the deformations of the nuclei  decrease  and
the values of the corresponding Coulomb barriers  increase.
%On other side,
% the values of the Coulomb barriers decrease
% by about of 0.3--0.4 MeV with decreasing mass asymmetry.
%Since the former effect is stronger than last one,
As a result, the transfer
%weakly influences or even
suppresses the capture process in these reactions at the sub-barrier energies.
The suppression becomes  stronger with decreasing  energy  (Figs.~7-9).
As seen in Fig.~7,
the capture cross sections calculated
 without two-neutron transfer are larger than those  calculated with two-neutron transfer in the case
 of
the $^{32}$S+$^{110}$Pd reaction.
The enhancement of the sub-barrier fusion  for
the reactions with $^{32}$S with respect to the reactions with $^{36}$S is  related to a larger deformation
of $^{34}$S
in comparison with $^{36}$S. We  observe the same behavior in the reactions  $^{32,36}$S+$^{94,96,98,100}$Mo.

%In Figs.~12 and 13, we observe the same behavior in the reactions
%$^{32}$S($\beta_2=0.312$)+$^{100}$Mo($\beta_2=0.231$)$\to ^{34}$S($\beta_2=0.252$)+$^{98}$Mo($\beta_2=0.168$),
%$^{32}$S($\beta_2=0.312$)+$^{98}$Mo($\beta_2=0.168$)$\to ^{34}$S($\beta_2=0.252$)+$^{96}$Mo($\beta_2=0.172$),
%$^{32}$S($\beta_2=0.312$)+$^{96}$Mo($\beta_2=0.172$)$\to ^{34}$S($\beta_2=0.252$)+$^{94}$Mo($\beta_2=0.151$),
%$^{32}$S($\beta_2=0.312$)+$^{94}$Mo($\beta_2=0.151$)$\to ^{34}$S($\beta_2=0.252$)+$^{92}$Mo($\beta_2=0$).
%In the reactions with $^{36}$S the Coulomb barriers are about 1.2 MeV lower than those with $^{32}$S.
%However, at the same $E_{\rm c.m.}/V_b< 0.95$
%the cross sections  in the reactions with $^{32}$S
% are always larger (Figs.~12 and 13) because of the deformation effects.
% In the case of $^{36}$S the cross section decreases steeper under the barrier for all target isotopes of Mo
%with exception of  $^{92}$Mo. For $^{92}$Mo, the cross section is larger in the
%reaction with $^{36}$S ($Q_{2n}>0$) than in reaction with $^{32}$S ($Q_{2n}<0$).
%One can see that after the $2n$-transfer in the reaction  (Fig.~13)
%$^{36}$S($\beta_2=0$)+$^{92}$Mo($\beta_2=0$)$\to ^{34}$S($\beta_2=0.252$)+$^{94}$Mo($\beta_2=0.151$)
%the deformation of the target-like nucleus becomes larger than
%in the reaction $^{32}$S($\beta_2=0.312$)+$^{92}$Mo($\beta_2=0$).
%As a result, the cross section decreases steeper under the barrier
%in the case of the $^{32}$S+$^{92}$Mo reaction.

%
\begin{figure}[htb]
\centering
%\sidecaption
\includegraphics[scale=1]{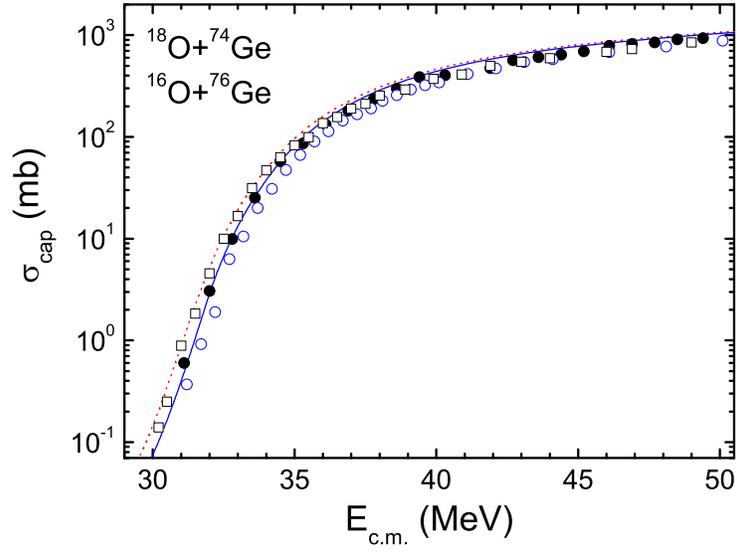}
\caption{
(Color online)
The calculated (solid line) capture cross sections vs $E_{\rm c.m.}$ for the reactions
$^{16}$O+$^{76}$Ge  and $^{18}$O+$^{74}$Ge (the curves coincide).
For the $^{18}$O+$^{74}$Ge reaction,
the calculated capture cross sections without
 neutron transfer are shown by dotted line.
The experimental data  for the reactions $^{16}$O+$^{76}$Ge (open circles) and $^{18}$O+$^{74}$Ge (open squares)
are from Ref.~\protect\cite{Jia}.
The experimental data  for the $^{16}$O+$^{76}$Ge reaction (solid circles)
are from Ref.~\protect\cite{16OAGe}.
}
\label{10_fig}
\end{figure}
\begin{figure}[htb]
\centering
%\sidecaption
\includegraphics[scale=1]{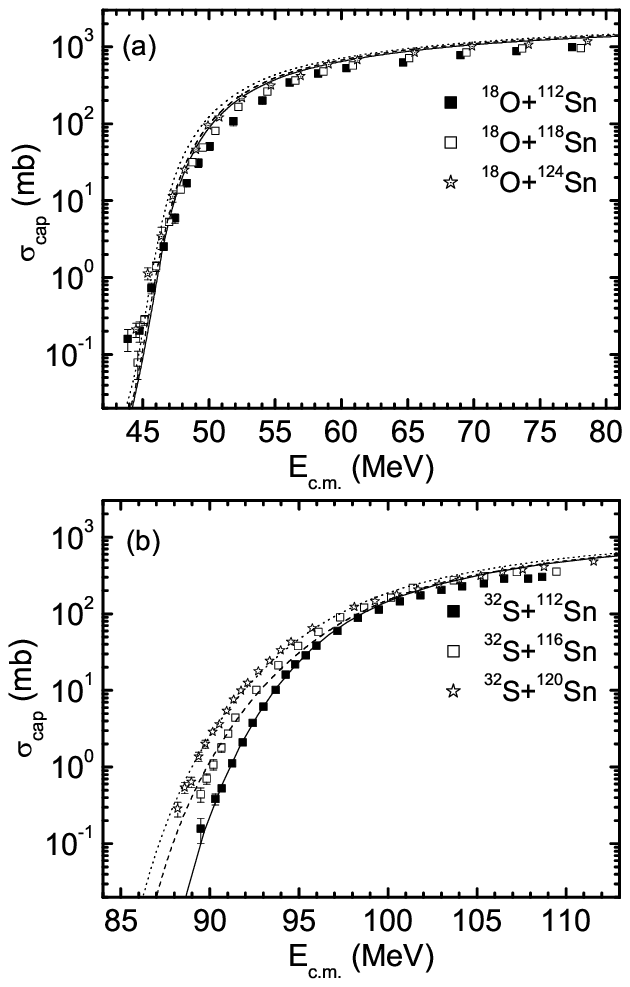}
\caption{
The calculated capture cross sections vs $E_{\rm c.m.}$ for the  reactions
$^{18}$S+$^{112,118,124}$Sn (solid, dashed and dotted lines, respectively) (a)
and
$^{32}$S+$^{112,116,120}$Sn (solid, dashed and dotted lines, respectively) (b).
The experimental data (symbols) are  from Ref.~\protect\cite{AOASn,Tripathi}.
}
\label{11_fig}
\end{figure}
Figures 10 and 11 show  the excitation functions  for the reactions
$^{18}$O+$^{74}$Ge,$^{112,118,124}$Sn and $^{32}$S+$^{112,116}$Sn.
For the $^{32}$S-induced reactions, $Q_{2n}>0$.
For the projectile
%$^{16}$O there are no neutron
%transfer channels with positive $Q$-value, but for
$^{18}$O there is a large range of positive  $Q_{2n}$-values, for example,
varying from 1.4 MeV for $^{18}$O+$^{124}$Sn up to 5.5 MeV for $^{18}$O+$^{112}$Sn.
The agreement between the
calculated results and the experimental data~\cite{Jia,AOASn} is rather good.
As seen in Fig.~11, the cross sections
increase systematically with the target mass number  and run
nearly similarly down to the lowest energy treated.
In the reactions
$^{32}$S($\beta_2=0.312$)+$^{112}$Sn($\beta_2=0.123$)$\to ^{34}$S($\beta_2=0.252$)+$^{110}$Sn($\beta_2=0.122$),
$^{32}$S($\beta_2=0.312$)+$^{116}$Sn($\beta_2=0.112$)$\to ^{34}$S($\beta_2=0.252$)+$^{114}$Sn($\beta_2=0.121$),
$^{18}$O($\beta_2=0.1$) + $^{74}$Ge($\beta_2=0.283$)$\to ^{16}$O($\beta_2=0$) + $^{76}$Ge($\beta_2=0.262$),
$^{18}$O($\beta_2=0.1$)+$^{112}$Sn($\beta_2=0.123$)$\to ^{16}$O($\beta_2=0$)+$^{114}$Sn($\beta_2=0.121$),
$^{18}$O($\beta_2=0.1$)+$^{118}$Sn($\beta_2=0.111$)$\to ^{16}$O($\beta_2=0$)+$^{120}$Sn($\beta_2=0.104$),
and
$^{18}$O($\beta_2=0.1$)+$^{124}$Sn($\beta_2=0.095$)$\to ^{16}$O($\beta_2=0$)+$^{126}$Sn($\beta_2=0.09$)
the $2n$-transfer suppresses the capture process  (Figs. 10 and 11).
The sub-barrier capture cross sections
for the systems $^{18}$O+$^{A}$Sn studied here do not show any strong
dependence on the mass number of the target isotope.
%
%Figures  4-7 show the capture excitation function  for the reactions
%$^{16,18}$O+$^{76,74}$Ge, and
%$^{16,18}$O+$^{114,112,120,118,126,124}$Sn
%as a function of  bombarding energy.
%One can see a rather good agreement between the
%calculated results and the experimental data~\cite{Jia,16OAGe,AOASn}
%for the reactions
%$^{16}$O+$^{76}$Ge,
% and
%$^{18}$O+$^{112,118,124}$Sn.
%The $Q_{2n}$-values for the $2n$-transfer
%processes are positive (negative) for all reactions with $^{18}$O ($^{16}$O).
%Thus, the neutron transfer can be important for the reactions with  the $^{16}$O beam.
Our results show that cross sections for reactions
$^{16}$O+$^{76}$Ge ($^{16}$O+$^{114,120,126}$Sn) [$Q_{2n}<0$]
and $^{18}$O+$^{74}$Ge ($^{18}$O+$^{112,118,124}$Sn) are very similar (Fig. 10).
%The reason of such behavior is that after the $2n$-transfer in the system
%$^{18}$O+$^{A-2}$X$\to ^{16}$O+$^{A}$X the deformations
%remain to be  similar.
%As a result, the corresponding Coulomb barriers of the  systems
%$^{18}$O+$^{A-2}$X and $^{16}$O+$^{A}$X
% are almost the same and, correspondingly, their capture cross
%sections  coincide.
Just the same behavior
was observed in the recent experiments $^{16,18}$O+$^{76,74}$Ge~\cite{Jia}.

%In the reaction
%$^{32}$S($\beta_2=0.312$)+$^{132}$Sn($\beta_2=0$)$\to ^{34}$S($\beta_2=0.252$)+$^{130}$Sn($\beta_2=0$)
%($Q_{2n}=6.5$ MeV)
%we expect a  suppression of the capture cross section due to the neutron transfer
%leading to
%$\beta_2(^{32}$S$)>\beta_2(^{34}$S$)$~(Fig.~15). As seen in Fig. 15,
%in the $^{32}$S+$^{132}$Sn reaction
% the cross section at sub-barrier energies is enhanced in comparison with
% the
%$^{36}$S($\beta_2=0$)+$^{132}$Sn$(\beta_2=0$) reaction
%which has a non-positive $Q$-value for the neutron transfer channel.
%This is related to the deformation effect: the deformation
%of $^{34}$S in the first 2$^+$ state
%is much larger than  that of $^{36}$S in the ground state.

\subsection{Neutron transfer in reactions with weakly bound nuclei}
\label{sec-4}
After the neutron transfer in the  reactions
$^{13}$C+$^{232}$Th($\beta_2=0.261$)$\to ^{14}$C($\beta_2=-0.36$)+$^{231}$Th($\beta_2=0.261$) ($Q_{1n}=1.74$ MeV),
%$^{14}$C+$^{232}$Th($\beta_2=0.261$) ($Q_{1n,2n}<0$),\\
$^{15}$C+$^{232}$Th($\beta_2=0.261$)$\to ^{14}$C($\beta_2=-0.36$)+$^{233}$Th($\beta_2=0.261$) ($Q_{1n}=3.57$ MeV)
the deformations of the target or projectile nuclei
in these reactions and in the $^{14}$C+$^{232}$Th($\beta_2=0.261$) ($Q_{1n,2n}<0$) reaction are the same.
%It means that the slopes of the excitation functions are the same.
In Fig.~12 the calculated cross sections  slightly
increase with the mass number of C, and are
nearly parallel down to the lowest energy treated.
There is a relatively good agreement between the
calculated results \cite{EPJSub3} and the experimental data \cite{Alcorta,CTh}
for the reactions $^{12,13,14}$C+$^{232}$Th,
but the experimental enhancement of the cross section in the
$^{15}$C+$^{232}$Th reaction  at sub-barrier energies cannot be explained  with
 our and other \cite{Alcorta} models.
Because we take into account the neutron transfer
($^{15}$C$\to ^{14}$C), one can suppose that this discrepancy is attributed to the
influence of the breakup channel \cite{Gomes}
which is not considered in our model. However, it is unclear why the
breakup process influences only two experimental points at lowest energies.
Different deviations of these points in energy from the
calculated curve in Fig.~12 create doubt in an influence of the breakup on the
kinetic energy.
So, additional experimental and theoretical
investigations are desirable.
\begin{figure}[htb]
\centering
%\sidecaption
\includegraphics[scale=1]{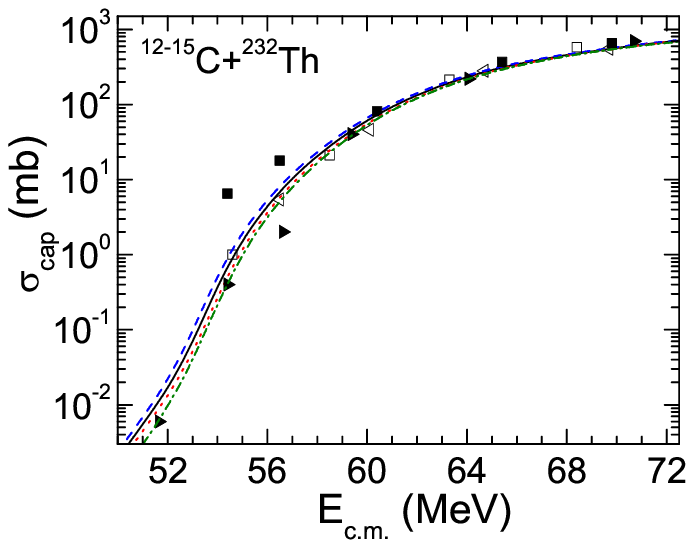}
\caption{
(Color online)
The calculated (lines) and experimental (symbols)
capture cross sections
vs $E_{\rm c.m.}$  for the  reactions
$^{12}$C+$^{232}$Th   (dash-dotted line, solid triangles),
$^{13}$C+$^{232}$Th (dotted line,  open  triangles),
$^{14}$C+$^{232}$Th (solid line, open squares), and
$^{15}$C+$^{232}$Th (dashed line,  solid squares).
The experimental  data   are  from Refs.~\protect\cite{Alcorta,CTh}.
}
\label{12_fig}
\end{figure}
\begin{figure}[htb]
\centering
%\sidecaption
\includegraphics[scale=1]{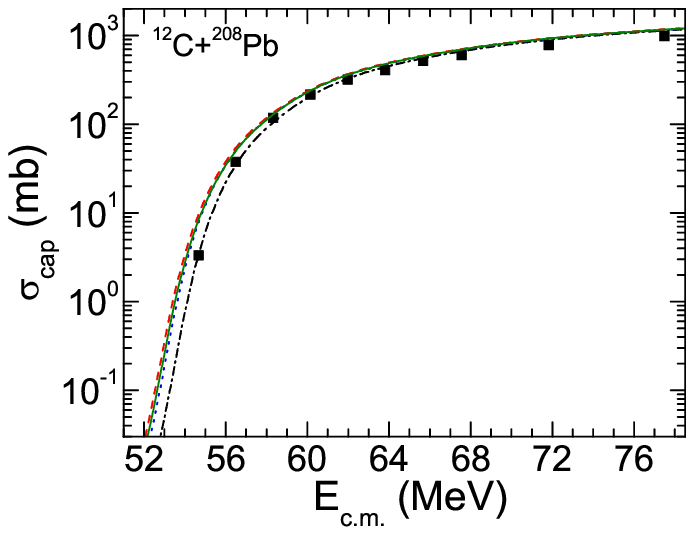}
\caption{
The calculated (lines)
and experimental
(symbols) capture cross sections
vs $E_{\rm c.m.}$
for the  reactions
$^{12}$C+$^{208}$Pb (dash-dotted line),
$^{13}$C+$^{208}$Pb (dotted line),
$^{14}$C+$^{208}$Pb (solid line),
and
$^{15}$C+$^{208}$Pb (dashed line).
The experimental  data  (solid squares) for the
$^{12}$C+$^{208}$Pb
reaction are  from Ref.~\protect\cite{12C208Pb}.
}
\label{13_fig}
\end{figure}

The question is whether the
fusion of nuclei involving weakly bound neutrons is enhanced
or suppressed at low energies. This question can been addressed
to the systems $^{12-15}$C+$^{208}$Pb~\cite{Alamanos3}.
After the neutron
transfer in the  reactions
$^{13}$C+$^{208}$Pb($\beta_2=0$)$\to ^{14}$C($\beta_2=-0.36$)+$^{207}$Pb($\beta_2=0$) ($Q_{1n}=1.74$ MeV),
%$^{14}$C+$^{208}$Pb($\beta_2=0$) ($Q_{1n,2n}<0$),\\
$^{15}$C+$^{208}$Pb($\beta_2=0$)$\to ^{14}$C($\beta_2=-0.36$)+$^{209}$Pb($\beta_2=0.055$) ($Q_{1n}=3.57$ MeV)
the deformations of the light nuclei
are the same as in the $^{14}$C+$^{208}$Pb($\beta_2=0$) ($Q_{1n,2n}<0$) reaction. The heavy nuclei
are almost spherical.
This means that the slopes of the excitation functions
are  almost the same (Fig.~13). As in the case of the $^{15}$C+$^{232}$Th reaction,
we do not expect
 enhancement of the capture cross section  in  the  $^{15}$C+$^{208}$Pb
 reaction owing to the neutron transfer.
The same effect was observed in Ref.~\cite{Alamanos3}.
The study of the reactions $^{15}$C+$^{208}$Pb,$^{232}$Th at sub-barrier energies
provides a good test for the verification  of the effect of
 weakly bound nuclei on fusion and capture
 because it reveals the role of other effects besides neutron transfer.
\begin{figure}[htb]
\centering
%\sidecaption
\includegraphics[scale=1]{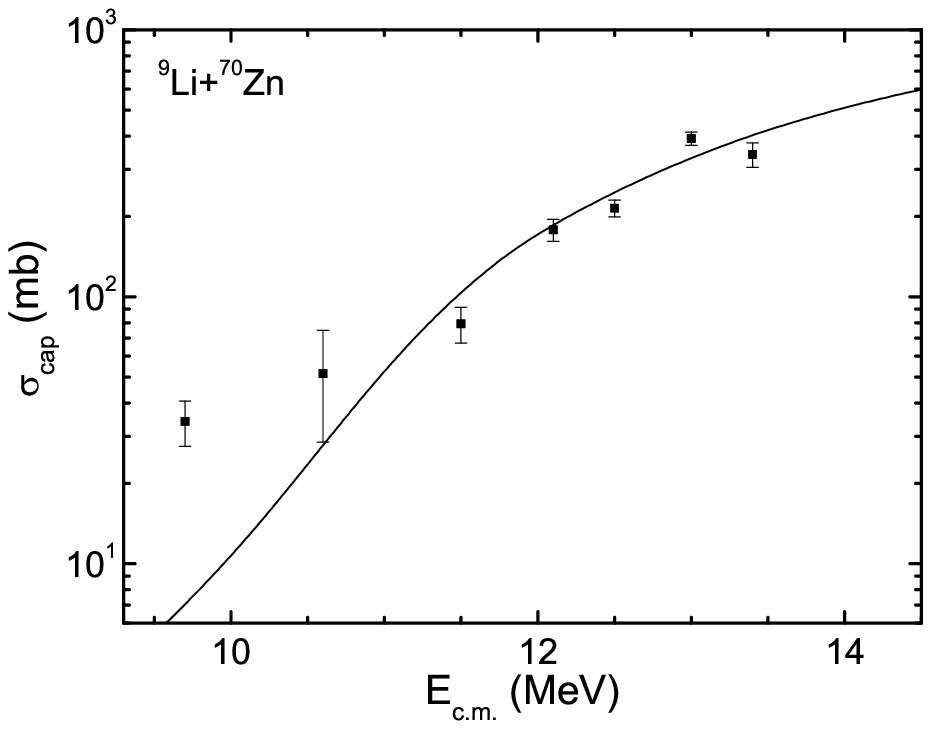}
\caption{
The calculated (solid line) and experimental (symbols)
capture cross sections  vs $E_{\rm c.m.}$ for the reaction
$^{9}$Li+$^{70}$Zn.
The experimental data
are from Ref.~\protect\cite{Vino}.
}
\label{14_fig}
\end{figure}

By assuming that
the $2n$-transfer process takes place and
the break-up channels are closed,
one can predict almost the same capture cross sections
for the reaction with large positive $Q_{2n}$ value
 $^{6}$He+$^{206}$Pb
($^{9}$Li+$^{68}$Zn) and for the complemented reaction
 $^{4}$He+$^{208}$Pb
($^{7}$Li+$^{70}$Zn).
%Here, the break-up channels are not taken into consideration in the calculations.
Indeed, after the transfer in the reactions
$^{6}$He+$^{206}$Pb$\to ^{4}$He($\beta_2=0$)+$^{208}$Pb($\beta_2=0.055$)   ($Q_{2n}=13.13$ MeV),
$^{9}$Li+$^{86}$Zn$\to ^{7}$Li($\beta_2\approx 0.4$)+$^{70}$Zn($\beta_2=0.248$) ($Q_{2n}=9.60$ MeV)
they become equivalent to the reactions  $^{4}$He+$^{208}$Pb and $^{7}$Li+$^{70}$Zn.
Therefore, the slopes of the excitation functions
in the reactions with $^{6}$He ($^{9}$Li) and  $^{4}$He ($^{7}$Li) should be  similar.
This conclusion supports  the experimental data of Ref.~\cite{Wolski},
where the authors concluded that the fusion enhancement in the $^{6}$He+$^{206}$Pb reaction
(with respect to the $^{4}$He+$^{208}$Pb reaction)
is rather small or absent.

By assuming that
the $2n$-transfer process occurs, we calculated the capture cross sections for the
$^{9}$Li+$^{70}$Zn reaction (Fig.~14). The agreement with the experimental data of Ref.~\cite{Vino}
is quite satisfactory. At
 lowest energies, the calculated cross section is by factor of $\sim 5$ less than the experimental value.
% At the lowest sub-barrier energy a difference by a factor of $\sim 5$ is observed.
The experimental data are well reproduced by the model~\cite{Bala} where
two-neutron transfer from the $^{70}$Zn leads to $^{11}$Li halo structure and molecular
bond between the nuclei in contact enhances the fusion cross section.
Note that two-neutron transfer $^{9}$Li+$^{70}$Zn$\to ^{7}$Li+$^{72}$Zn with $Q_{2n}=8.6$ MeV is much energetically
favorable than the two-neutron transfer $^{9}$Li+$^{70}$Zn$\to ^{11}$Li+$^{68}$Zn with $Q_{2n}=-15.4$ MeV.
These observations deserve further experimental and theoretical
investigations including the breakup channel.

\subsection{Breakup probabilities}
\label{sec-5}
%We try to reveal a systematic behavior of the complete fusion suppression
%as a function of the target charge $Z_T$ and colliding energy $E_{\rm c.m.}$
%by using the quantum diffusion approach~\cite{EPJSub4}
%and by comparing the calculated capture cross sections in the absence of breakup
%with the  experimental complete  fusion cross sections.
%The effects of deformation and neutron transfer on the complete fusion
% are taken into consideration in the calculations.
%
The difference between the calculated  capture cross section $\sigma_{cap}^{th}$ in the absence of breakup and
the experimental complete fusion cross section $\sigma_{fus}^{exp}$ can be ascribed
to the breakup effect with the  probability \cite{EPJSub4}
%Comparing $\sigma_{cap}^{th}$ and $\sigma_{fus}^{exp}$, one can estimate the breakup probability
\begin{eqnarray}
P_{\rm BU}=1-\sigma_{fus}^{exp}/\sigma_c^{th}.
\end{eqnarray}
If at some energy $\sigma_{fus}^{exp}>\sigma_{cap}^{th}$, the values
of $\sigma_{cap}^{th}$ was normalized so to have $P_{\rm BU}\ge 0$ at any energy.
Note that $\sigma_{fus}^{exp}=\sigma_{fus}^{noBU}+\sigma_{fus}^{BU}$
contains the contribution from two processes: the direct fusion of the
projectile with the target ($\sigma_{fus}^{noBU}$), and the breakup of the
projectile followed by the fusion of the two projectile fragments with
the target ($\sigma_{fus}^{BU}$). A more adequate estimate of the breakup
probability would then be:
\begin{eqnarray}
P_{\rm BU}=1-\sigma_{fus}^{noBU}/\sigma_{cap}^{th},
\end{eqnarray}
which
leads to larger values of $P_{\rm BU}$ than the expression employed by us.
However, the ratio between $\sigma_{fus}^{noBU}$ and $\sigma_{fus}^{BU}$
cannot be measured experimentally  but can be estimated with the approach
suggested in Ref.~\cite{Maximka}. The parameters of the potential
are taken to fit the height of the Coulomb barrier obtained in our calculations.
The parameters of the breakup function \cite{Maximka} are set to describe the value of
$\sigma_{fus}^{exp}$. As shown in Ref.~\cite{Maximka} and in our calculations, in
the $^8$Be+$^{208}$Pb reaction the fraction of
$\sigma_{fus}^{BU}$ in $\sigma_{fus}^{exp}$ does not exceed few percents
at $E_{\rm c.m.}-V_b<$4 MeV. This fraction rapidly increases and reaches about 12--20\%, depending on the reaction, at
$E_{\rm c.m.}-V_b\approx$10 MeV. Because we are mainly interested in
the energies near and below the barrier, the estimated $\sigma_{fus}^{BU}$ does not exceed 20\% of
$\sigma_{fus}^{exp}$ at $E_{\rm c.m.}-V_b <$10 MeV. The results for $P_{\rm BU}$ are presented,
taking $\sigma_{fus}^{noBU}$ into account in Eq.~(4).
\begin{figure}[htb]
\centering
%\sidecaption
\includegraphics[scale=1]{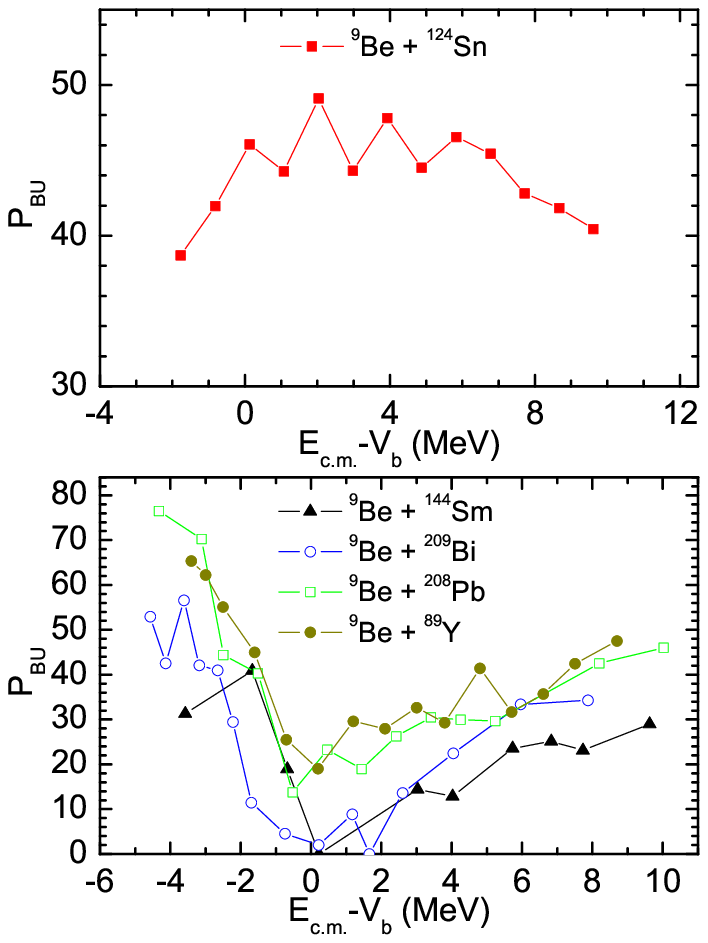}
\caption{
(Color online) The dependence of the extracted breakup probability $P_{BU}$
vs $E_{c.m.}-V_b$ for the indicated reactions with $^{9}$Be-projectiles in \%.
Formula (4) was used.
}
\label{15_fig}
\end{figure}
\begin{figure}[htb]
\centering
%\sidecaption
\includegraphics[scale=1]{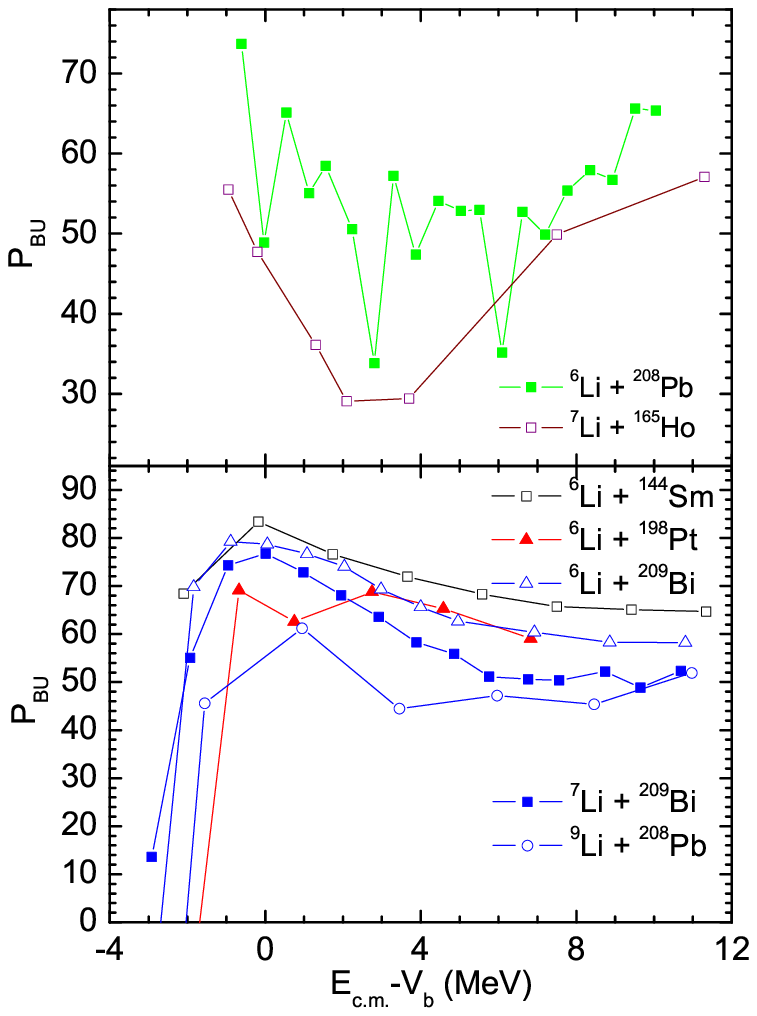}
\caption{
(Color online) The same as in Fig.~15, but for the indicated reactions with $^{6,7,9}$Li-projectiles.
}
\label{16_fig}
\end{figure}

As seen in Figs.~15 and 16, at energies above the Coulomb barriers the values of $P_{\rm BU}$ vary from 0 to 84\%.
In the reactions $^9$Be+$^{144}$Sm,$^{208}$Pb,$^{209}$Bi the value of $P_{\rm BU}$ increases with charge
number of the target at $E_{\rm c.m.}-V_b>3$ MeV.
This was also noted in Ref.~\cite{PRSGomes5}. However, the reactions $^9$Be+$^{89}$Y,$^{124}$Sn are out of this
systematics. In the reactions $^6$Li+$^{144}$Sm,$^{198}$Pt,$^{209}$Bi the value of $P_{\rm BU}$ decreases with
increasing charge number of the target at $E_{\rm c.m.}-V_b>3$ MeV. While in the reactions
$^9$Be+$^{89}$Y,$^{144}$Sm,$^{208}$Pb,$^{209}$Bi the value of  $P_{\rm BU}$ has a minimum
at $E_{\rm c.m.}-V_b\approx 0$ and a maximum at $E_{\rm c.m.}-V_b\approx -(1-3)$ MeV, in the
$^9$Be+$^{124}$Sn reaction the value of $P_{\rm BU}$ steadily decreases with energy. In the reactions
$^6$Li+$^{144}$Sm,$^{198}$Pt,$^{209}$Bi, $^7$Li+$^{208}$Pb,$^{209}$Bi, and $^9$Li+$^{208}$Pb there is maximum of
$P_{\rm BU}$
at $E_{\rm c.m.}-V_b\approx -(0-1)$ MeV. However, in the reactions $^6$Li+$^{208}$Pb and $^7$Li+$^{165}$Ho $P_{\rm BU}$
has a minima $E_{\rm c.m.}-V_b\approx 2$ MeV and no maxima at $E_{\rm c.m.}-V_b\approx 0$.
For $^9$Be, the breakup threshold is slightly larger than for $^6$Li.
Therefore, we cannot explain a larger breakup probability
at smaller $E_{\rm c.m.}-V_b$ in the case of $^9$Be.

\section{Quasi-elastic and elastic backscattering - tools for search of  breakup process
in reactions with weakly bound projectiles}
\label{sec-6}
The lack of a clear systematic behavior of the complete fusion suppression
as a function of the target charge  requires new additional
experimental and theoretical studies.
The quasi-elastic backscattering
has been used~\cite{Timmers,EPJSub4}
as an alternative to investigate fusion (capture) barrier
distributions, since this process is complementary to fusion.
%The sum of elastic and inelastic scattering is much easier to
%measure than fusion, and its barrier distribution is derived by
%the first derivative of its relative cross section to the Rutherford
%cross section with respect to the center-of-mass energy.
Since
the quasi-elastic experiment is usually not as complex as the capture (fusion) and breakup
measurements, they are well suited to survey the breakup probability.
There is a direct relationship between the capture,
the quasi-elastic scattering  and the breakup processes, since any loss from
the quasi-elastic and breakup channel contributes directly to capture (the conservation of the total reaction flux):
\begin{eqnarray}
P_{qe}(E_{\rm c.m.},J)+P_{cap}(E_{\rm c.m.},J)+P_{BU}(E_{\rm c.m.},J)=1,
\end{eqnarray}
where
$P_{qe}$  is the reflection quasi-elastic probability, $P_{BU}$ is the breakup  probability, and
$P_{cap}$ is the capture  probability.
The quasi-elastic scattering ($P_{qe}$) is the sum of
all direct reactions, which include elastic ($P_{el}$),
inelastic ($P_{in}$), and
a few nucleon transfer ($P_{tr}$) processes.
In Eq. (5)  we neglect the deep inelastic
collision process, since we are concerned with low energies.
%
%It has been proposed that to obtain
%the interaction barrier, quasielastic scattering should be measured
%at backward angles of nearly 180 degrees, where head-on collision
%is dominant~\cite{Timmers,Zhang,Sinha,Piasecki}.
%
Equation~(5) can be rewritten as
\begin{eqnarray}
\frac{P_{qe}(E_{\rm c.m.},J)}{1-P_{BU}(E_{\rm c.m.},J)}+\frac{P_{cap}(E_{\rm c.m.},J)}{1-P_{BU}(E_{\rm c.m.},J)}
=P_{qe}^{noBU}(E_{\rm c.m.},J)+P_{cap}^{noBU}(E_{\rm c.m.},J)=1,
\end{eqnarray}
where
$$P_{qe}^{noBU}(E_{\rm c.m.},J)=\frac{P_{qe}(E_{\rm c.m.},J)}{1-P_{BU}(E_{\rm c.m.},J)}$$
and
$$P_{cap}^{noBU}(E_{\rm c.m.},J)=\frac{P_{cap}(E_{\rm c.m.},J)}{1-P_{BU}(E_{\rm c.m.},J)}$$
are the quasi-elastic and  capture  probabilities, respectively,  in the absence of the
breakup process. From these expressions we obtain the useful formulas
\begin{eqnarray}
\frac{P_{qe}(E_{\rm c.m.},J)}{P_{cap}(E_{\rm c.m.},J)}=\frac{P_{qe}^{noBU}(E_{\rm c.m.},J)}{P_{cap}^{noBU}(E_{\rm c.m.},J)}
=\frac{P_{qe}^{noBU}(E_{\rm c.m.},J)}{1-P_{qe}^{noBU}(E_{\rm c.m.},J)}=a.
\end{eqnarray}
Using  Eqs.~(5) and (7), we obtain the relationship between breakup and quasi-elastic processes:
\begin{eqnarray}
P_{BU}(E_{\rm c.m.},J)=1-P_{qe}(E_{\rm c.m.},J)[1+1/a]=1-\frac{P_{qe}(E_{\rm c.m.},J)}{P_{qe}^{noBU}(E_{\rm c.m.},J)}.
\end{eqnarray}
%The last equation is one of important results of the present paper.
%Analogously one can find other expression
%\begin{eqnarray}
%P_{BU}(E_{\rm c.m.},J)=
%1-P_{cap}(E_{\rm c.m.},J)/P_{cap}^{noBU}(E_{\rm c.m.},J),
%\end{eqnarray}
%which relates the breakup and capture processes.

The reflection quasi-elastic probability
%\begin{eqnarray}
$P_{qe}(E_{\rm c.m.},J=0)=d\sigma_{qe}/d\sigma_{Ru}$
%\end{eqnarray}
for bombarding energy
$E_{\rm c.m.}$ and
angular momentum $J=0$ is given by the ratio of
the quasi-elastic differential cross section $\sigma_{qe}$  and
Rutherford differential cross section $\sigma_{Ru}$
at 180 degrees~\cite{Timmers}.
Employing Eq.~(8) and the  experimental quasi-elastic backscattering data
with toughly
and
weakly bound isotopes-projectiles and the same compound nucleus,
one can extract the breakup probability of the exotic nucleus.
For example, using Eq.~(8)  at backward angle,
 the experimental $P_{qe}^{noBU}$[$^{4}$He+$^{A}$X] of the
$^{4}$He+$^{A}$X reaction with toughly bound nuclei (without breakup), and $P_{qe}$[$^{6}$He+$^{A-2}$X]
of the $^{6}$He+$^{A-2}$X reaction with weakly bound projectile (with breakup), and taking into consideration
$V_b$($^{4}$He+$^{A}$X)$\approx V_b$($^{6}$He+$^{A-2}$X)
for the very asymmetric systems,
one can extract the breakup probability of the $^{6}$He:
\begin{eqnarray}
P_{BU}(E_{\rm c.m.},J=0)=
1-\frac{P_{qe}(E_{\rm c.m.},J=0)[^{6}He+^{A-2}{\rm X}]}{P_{qe}^{noBU}(E_{\rm c.m.},J=0)[^{4}He+^{A}{\rm X}]}.
\end{eqnarray}
Comparing the experimental quasi-elastic backscattering
cross sections in the presence and absence  of breakup
data in the reaction pairs
 $^{6}$He+$^{68}$Zn    and    $^{4}$He+$^{70}$Zn,
 $^{6}$He+$^{122}$Sn   and    $^{4}$He+$^{124}$Sn,
 $^{6}$He+$^{236}$U    and    $^{4}$He+$^{238}$U,
 $^{8}$He+$^{204}$Pb   and    $^{4}$He+$^{208}$Pb,
 $^{8}$Li+$^{207}$Pb   and    $^{7}$Li+$^{208}$Pb,
 $^{7}$Be+$^{207}$Pb   and   $^{10}$Be+$^{204}$Pb,
 $^{9}$Be+$^{208}$Pb   and   $^{10}$Be+$^{207}$Pb,
$^{11}$Be+$^{206}$Pb   and   $^{10}$Be+$^{207}$Pb,
 $^{8}$B+$^{208}$Pb    and   $^{10}$B+$^{206}$Pb,
 $^{8}$B+$^{207}$Pb    and   $^{11}$B+$^{204}$Pb,
 $^{9}$B+$^{208}$Pb    and   $^{11}$B+$^{206}$Pb,
$^{15}$C+$^{204}$Pb   and    $^{12}$C+$^{207}$Pb,
$^{15}$C+$^{206}$Pb   and    $^{13}$C+$^{208}$Pb,
$^{15}$C+$^{207}$Pb   and    $^{14}$C+$^{208}$Pb,
$^{17}$F+$^{206}$Pb   and    $^{19}$F+$^{208}$Pb,
leading to the same corresponding compound nuclei,
one can analyze  the role of the breakup channels
in the reactions with
the light weakly bound projectiles $^{6,8}$He, $^{8}$Li, $^{7,9,11}$Be,
 $^{8,9}$B, $^{15}$C, and  $^{17}$F  at near and above the barrier energies.
On other side,  the experimental uncertainties could be probably smaller when
the same target-nucleus $^{A}$X  is used in the
reactions with weakly and toughly bound isotopes.
Then, one can extract the breakup probability of the $^{6}$He
[$\Delta E=V_b(^{4}{\rm He}+^{A}{\rm X}) - V_b(^{6}{\rm He}+^{A}{\rm X})]$:
\begin{eqnarray}
P_{BU}(E_{\rm c.m.},J=0)=
1-\frac{P_{qe}(E_{\rm c.m.},J=0)[^{6}{\rm He}+^{A}{\rm X}]}{P_{qe}^{noBU}(E_{\rm c.m.}+\Delta E,J=0)[^{4}{\rm He}+^{A}{\rm X}]}.
\end{eqnarray}
For the very asymmetric systems, one can neglect  $\Delta E$.

Using the conservation of the total reaction flux,
analogously one can find the following expression
\begin{eqnarray}
P_{BU}(E_{\rm c.m.},J)=
1-\frac{P_{el}(E_{\rm c.m.},J)}{P_{el}^{noBU}(E_{\rm c.m.},J)},
\end{eqnarray}
which relates the breakup and elastic scattering  processes.
$P_{el}^{noBU}(E_{\rm c.m.},J)$ is the elastic scattering probability
in the absence of the breakup process. So,
one can extract the breakup probability of the $^{6}$He at the backward angle:
\begin{eqnarray}
P_{BU}(E_{\rm c.m.},J=0)=
1-\frac{P_{el}(E_{\rm c.m.},J=0)[^{6}{\rm He}+^{A-2}{\rm X}]}{P_{el}^{noBU}(E_{\rm c.m.},J=0)[^{4}{\rm He}+^{A}{\rm X}]}
\end{eqnarray}
or
\begin{eqnarray}
P_{BU}(E_{\rm c.m.},J=0)=
1-\frac{P_{el}(E_{\rm c.m.},J=0)[^{6}{\rm He}+^{A}{\rm X}]}{P_{el}^{noBU}(E_{\rm c.m.}+\Delta E,J=0)[^{4}{\rm He}+^{A}{\rm X}]}.
\end{eqnarray}
One concludes that
the quasi-elastic or elastic backscattering technique
could be a very important tool in breakup research.
We propose to extract the breakup probability
directly from the quasi-elastic or elastic backscattering
probabilities of systems mentioned above.

\section{Summary}
The quantum diffusion approach was applied to study
the role of the neutron transfer with   positive $Q$-value
in the capture  reactions  at sub-, near- and above-barrier energies.
We demonstrated   a good agreement of the theoretical calculations
with the  experimental data.
We found, that the change of the magnitude of the
 capture cross section after the neutron transfer
occurs due to the change of the deformations of  nuclei.
The  effect of the  neutron transfer is an indirect effect of the quadrupole
deformation.
When after the neutron transfer the deformations of nuclei
do not change  or slightly decrease,
the neutron transfer weakly influences or suppresses the capture cross section.
Good examples for this effect  are the capture reactions
$^{60}$Ni + $^{100}$Mo,$^{150}$Nd,
$^{18}$O +  $^{64}$Ni,$^{112,114,116,118,120,122,124}$Sn,$^{204,206}$Pb,
and
$^{32}$S+$^{96}$Zr,$^{94,96,98,100}$Mo,$^{100,102,104}$Ru,$^{104,106,108,110}$Pd,$^{112,114,116,118,120,122,124}$Sn.
at sub-barrier energies.
Thus, the general point of view
that  the sub-barrier capture
(fusion) cross section strongly
increases because of the neutron transfer with a  positive $Q$-values
has to be revised.

The neutron transfer effect can lead to a   weak influence of halo-nuclei on the capture.
Comparing  the capture cross sections  calculated  without the breakup effect
and experimental complete fusion cross sections, the breakup  was analyzed
in reactions with weakly bound projectiles. A trend of a systematic
behavior for the complete fusion suppression as a function of the target charge
and bombarding energy is not achieved.
The  quasi-elastic or elastic backscattering  was suggested to be an useful
tool to study  the behavior of the breakup probability.

%
%Within the quantum diffusion approach  is shown that in some cases the neutron transfer  with positive $Q$-value can weakly
%influence or suppress the sub-barrier capture through
%the change of the deformations of the colliding nuclei.
%Based on our  analysis a weak influence of halo-nuclei on the capture
%is demonstrated.
%A trend of a systematic
%behavior for the complete fusion suppression as a function of the target charge
%and bombarding energy is not achieved.
%The  quasi-elastic and elastic backscattering  is suggested to be an useful
%tool to study  the behavior of the breakup probability in reactions with weakly bound projectiles.

We thank P.R.S.~Gomes and A.~L\'epina-Szily for fruitful discussions and suggestions.
This work was supported by DFG, NSFC, RFBR, and JINR grants.
The IN2P3(France)-JINR(Dubna) and Polish - JINR(Dubna) Cooperation Programmes
are gratefully acknowledged.\newline
%
% BibTeX or Biber users please use (the style is already called in the class, ensure that the "woc.bst" style is in your local directory)
% \bibliography{name or your bibliography database}
%
% Non-BibTeX users please use
%

\end{document}